\documentclass[pre,aps,preprint,floatfix,showpacs]{revtex4}
\usepackage{amssymb}
\usepackage{amsmath}
\usepackage{bm}
\usepackage{graphicx}
\usepackage{graphics}
\usepackage{color}
\usepackage{subfigure}
\usepackage[linktocpage,colorlinks=true,linkcolor=blue,citecolor=blue]{hyperref}

\begin{document}

\title{Endogenous Quasicycles and Stochastic Coherence in a Closed Endemic Model}

\author{Somdeb Ghose}
\email{somdeb@imsc.res.in}
\author{R. Adhikari}
\email{rjoy@imsc.res.in}
\affiliation{The Institute of Mathematical Sciences, CIT Campus, Tharamani, Chennai-600113, India}

\date{\today}

\begin{abstract} 
We study the role of demographic fluctuations in typical endemics as exemplified by the stochastic SIRS model. The birth-death master equation of the model is simulated using exact numerics and analysed within the linear noise approximation. The endemic fixed point is unstable to internal demographic noise, and leads to sustained oscillations. This is ensured when the eigenvalues ($\lambda$) of the linearised drift matrix are complex, which in turn, is possible only if detailed balance is violated. In the oscillatory state, the phases decorrelate asymptotically, distinguishing such oscillations from those produced by external periodic forcing. These so-called quasicycles are of sufficient strength to be detected reliably only when the ratio $|Im(\lambda)/Re(\lambda)|$ is of order unity. The coherence or regularity of these oscillations show a maximum as a function of population size, an effect known variously as stochastic coherence or coherence resonance. We find that stochastic coherence can be simply understood as resulting from a non-monotonic variation of $|Im(\lambda)/Re(\lambda)|$ with population size. Thus, within the linear noise approximation, stochastic coherence can be predicted from a purely deterministic analysis. The non-normality of the linearised drift matrix, associated with the violation of detailed balance, leads to enhanced fluctuations in the population amplitudes.
\end{abstract}

\maketitle
 
\section{Introduction} 

Two and a half centuries ago, D. Bernoulli \cite{bernoulli1760} used a nonlinear ordinary differential equation to study the effect of cow-pox inoculation on the spread of smallpox. This was one of the earliest examples of the mathematical study of epidemics. This field of study continues to hold the interest of the scientific community especially in the light of recent outbreaks of viral pandemics like SARS and H1N1. Kermack and McKendrick in their seminal paper \cite{kermack1927} put forward the classic Susceptible-Infected-Recovered (SIR) model of the spread of epidemics which, like most early epidemic models, assumes a homogeneously mixed population. More recent work focuses on the geotemporal spread of epidemics, especially on model networks \cite{kuperman2001, pastor-satorras2001}. However, homogeneous mixing models still prove to be useful \cite{keeling2005} and have been used to study various outbreaks of diverse size, fatality and chronology. Examples range from the study of the plague in the village of Eyam in 1665-66 \cite{raggett1982} to the Bombay plague of 1905-06 \cite{kermack1927} and the influenza epidemic in an English boarding school \cite{murray2002}. 

Mathematical models like the SIR model are usually analysed deterministically and are only exactly valid when the size of the population under consideration is exceedingly large. Fluctuations due to finite population sizes or due to external causes can give rise to phenomena which cannot be captured by deterministic mean-field models and necessitates the use of stochastic models. Bartlett \cite{bartlett1956, bartlett1957} was one of the first to realise that a stochastic description was necessary to explain the periodic recurrence of measles, a phenomenon which could not be explained by deterministic models \cite{soper1929, wilson1945}. Bartlett formulated \cite{bartlett1956} a stochastic version of the SIR model to describe the periodic recurrence of measles. 

The mechanism for the generation of sustained oscillations in population dynamics has been analysed within the stochastic framework \cite{nisbet1976} which concentrates on external fluctuations as the noise source. However, finite-sized populations give rise to fluctuations  whose relative amplitude is of the order of the inverse of the square root of the size of the population \cite{schrodinger1992}. The role played by this internal noise, arising out of demographic stochasticity, in the generation of sustained oscillations has been studied in a prey-predator model using a master equation approach by McKane and Newman \cite{mckane2005}. They have used the expansion method due to van Kampen \cite{vanKampen2007} in their analysis, which provides a systematic way of deriving the phenomonological equations due to Bartlett \cite{bartlett1956}. Alonso \emph{et al.} \cite{alonso2007} used similar techniques in an open model of infectious diseases within the homogeneous mixing assumption, while Rozhnova and Nunes \cite{rozhnova2009} applied systematic expansion to a closed epidemic model on networks, using a pair approximation. The oscillations generated and sustained by internal noise are called \emph{endogenous resonant quasicycles} and are qualitatively different from stochastic oscillations forced by external periodicities which are \emph{exogenous} \cite{nisbet1982}. The quality or coherence of these oscillations are intuitively expected to vary monotonically with the size of the population or equivalently, the relative noise amplitude. However, it has been observed in various theoretical models including the Fitz Hugh-Nagumo \cite{pikovsky1997} and gene circuit models \cite{hilborn2008} that the regularity or coherence of oscillations is small for low and high noise amplitudes and reaches a maximum for an intermediate value. This phenomenon is called stochastic coherence or coherence resonance and has also been observed in optical laser experiments \cite{giacomelli2000}.

In this work we analyse the generation of quasicycles due to internal noise, as well as the non-trivial variation of the quality of oscillation with respect to population size, in a closed epidemic model under the homogeneous mixing assumption. The closed system is relevant in many epidemiological situations, for instance in boarding houses \cite{murray2002}, or island communities, where no inflows or outfluxes occur. Further, the conservation of populations, as implied by a closed system, allows one to deal with a lower-dimensional problem. We exploit this in a systematic manner and show how the master equation can be marginalised using the conservation constraint. The existence of an endemic fixed point allows a two-stage linearisation procedure to be carried out on the model. The linear noise approximation, followed by a further linearisation about the endemic fixed point, reduces the model to the standard multivariate Ornstein-Uhlenbeck (OU) form. Exploiting the linear and Gaussian character of the multivariate OU process then allows for stochastic behaviour to be predicted from the deterministic part of the dynamics, in a spirit similar to the Onsager regression method of equilibrium statistical mechanics. 

Below we review (section \ref{Deterministic analysis}) the deterministic analysis of the SIRS model, emphasising the behaviour around the stable endemic fixed point. Perturbations about this fixed point decay either monotonically or in a damped oscillatory fashion. Stochastic analysis (sections \ref{Stochastic analysis} and \ref{Quasicycles}), however, shows that demographic noise destabilises this endemic fixed point, generating and sustaining oscillations. We show the existence of stochastic coherence (section \ref{Stochastic coherence}) by analytical means and then confirm it numerically. We show from purely deterministic analysis that there is stochastic coherence if the absolute value of the ratio of the imaginary and real parts of the eigenvalues of the linearised drift matrix shows a maximum when scanned against noise amplitude. The position of this maximum gives the population size corresponding to stochastic coherence for the relevant parameter values. We also show that (section \ref{Non-equilibrium}) it is not possible to observe endogenous quasicycles unless detailed balance is violated. Finally we look at the non-normal aspect of the governing dynamics (section \ref{Non-normality}) and observe that the fluctuation amplitudes of the populations increase due to non-normality.

\section{SIRS linear deterministic analysis}\label{Deterministic analysis}

The classic SIR model for infectious diseases (S stands for susceptibles, I for infected and R for recovered) considers the population to be homogeneously mixed and constant in total number \cite{kermack1927}. The SIRS model is a variant of the SIR model where the recovered section of the population lose their immunity after a delay and become susceptible. The nonlinear ODE system of the form $\dot{\mathbf{n}} = \mathbf{f}(\mathbf{n})$, where $\mathbf{n} = \{S, I, R\}$, describing the SIRS model is constrained by the fixed population size $\Omega$ and is hence a closed system.
\begin{eqnarray}
 \dot{S} &=& \alpha R - \beta S I \nonumber \\
 \dot{I} &=& \beta S I - \gamma I \\
 \dot{R} &=& \gamma I - \alpha R \nonumber
\end{eqnarray}
The rate of infection is $\beta$, the rate of recovery is $\gamma$ while $\alpha$ is the rate of loss of immunity. The fixed point ($\mathbf{n}=\mathbf{n}^{\ast}$) is given by
\begin{equation}
(S^{\ast}, I^{\ast}, R^{\ast}) = \left[ \frac{\gamma}{\beta}, \frac{\alpha}{\beta} \left( \frac{\beta \Omega-\gamma}{\alpha+\gamma} \right),  \frac{\gamma}{\beta} \left( \frac{\beta \Omega-\gamma}{\alpha+\gamma} \right) \right]
\label{SIRS fixed points}
\end{equation}
The steady state with zero infected is not of interest in the present study. The fixed point is endemic with non-zero infected in the steady state ($I^{\ast} >0$) when the condition $\beta \Omega > \gamma$ is satisfied.

Since there is a constraint in the system, $S+I+R=\Omega$, the $3 \times 3$ system is effectively a $2 \times 2$ system with $R = \Omega -S -I$. 
\begin{eqnarray}
 \dot{S} &=& \alpha (\Omega-S-I) - \beta S I \\
 \dot{I} &=& \beta S I - \gamma I \nonumber
\end{eqnarray}

\begin{figure}
\includegraphics[scale=0.60]{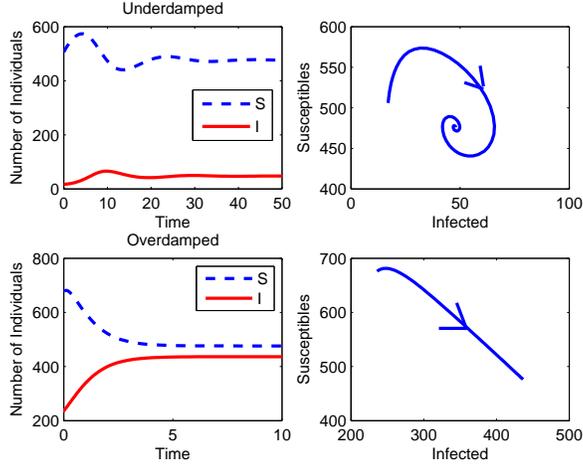}
\caption{\label{Over vs Underdamped decay}(Color online) Underdamped and overdamped decay of perturbations. The top two plots show underdamped decay with parameter values $\beta=0.0021,\alpha=0.1, \gamma=1.0$ and population size $\Omega = 1000$ (where the Jacobian has complex eigenvalues). The bottom two plots show overdamped decay with parameter values $\beta=0.0021,\alpha=5.0, \gamma=1.0$ and population size $\Omega = 1000$ (where the Jacobian has real eigenvalues). The S vs I plot for underdamped decay shows a spiral while that for overdamped decay does not. The former is a stable spiral while the latter is a stable node.}
\end{figure}

\begin{figure}
\includegraphics[scale=0.45]{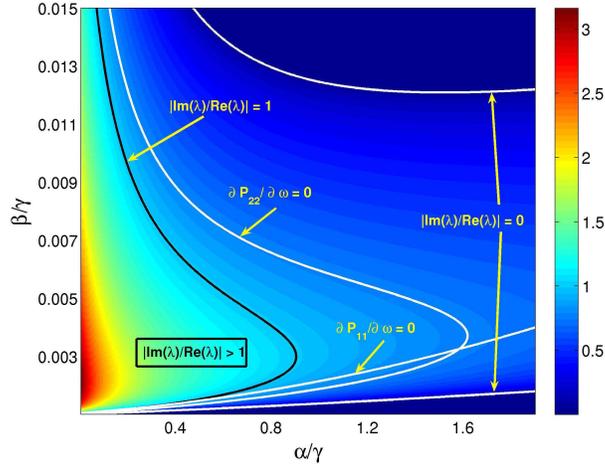}
\caption{\label{fig:ratio_Im_Re_Eig}(Color online) Plot of the absolute value of the ratio of the imaginary and real parts of the eigenvalues $\lambda$ of the linearised Jacobian matrix $\mathbf{A}$ against the dimensionless parameters $\beta/\gamma$ and $\alpha/\gamma$ and population size $\Omega = 1000$. The outermost white contours, labeled by $|{\rm Im}(\lambda)/{\rm Re}(\lambda)|= 0$, enclose the region where the eigenvalues are complex, which is a necessary condition for the existence of quasicycles. The imaginary parts of the eigenvalues are zero in the outermost two regions of the plot. The inner white contours, labeled ``$\partial P_{11}(\omega) / \partial \omega = 0$'' and ``$\partial P_{22}(\omega) / \partial \omega = 0$'', enclose the regions for which the PSD shows a peak. This is a sufficient condition for the existence of quasicycles. The innermost black contour, denoting $|{\rm Im}(\lambda)/{\rm Re}(\lambda)|= 1$ and marked as such, encloses the region where the quasicycles are of sufficient strength to be reliably detected. This region is labeled ``$|{\rm Im}(\lambda)/{\rm Re}(\lambda)| > 1$'', which is a necessary and sufficient condition for the reliable detection of quasicycles. Each condition presented above is stricter than the previous, leading to a nesting of regions of parameter space as regards the existence and detection of quasicycles.}
\end{figure}

The dynamics of small perturbations, $\delta \mathbf{n} = \{\delta S, \delta I\}$, about the fixed point are described by the linear ODE system $\delta \dot{\mathbf{n}} = \mathbf{A} \cdot \delta \mathbf{n}$. Here $A_{ij} = \partial f_i/\partial n_j |_{\mathbf{n} = \mathbf{n}^{\ast}} $ is the Jacobian matrix at the fixed point and is given by
\begin{equation}
 \mathbf{A} =
 \begin{bmatrix}
  - \alpha \left( \frac{\alpha+\beta \Omega}{\alpha + \gamma} \right) & -(\alpha+\gamma) \\
  \alpha \left( \frac{\beta \Omega -\gamma}{\alpha + \gamma} \right) & 0
 \end{bmatrix}
 \label{SIRS Linearised A matrix}
\end{equation}
Its eigenvalues are 
\begin{equation}
\lambda_{\pm} = \frac{1}{2} \left[ - \alpha \left( \frac{\alpha+\beta \Omega}{\alpha + \gamma} \right) \right. \left. \pm \sqrt{ \alpha^2 \left( \frac{\alpha+\beta \Omega}{\alpha + \gamma} \right)^2 - 4\alpha(\beta \Omega -\gamma)}  \right] 
\label{Jacobian eigenvalues}
\end{equation}
the real parts of which are always negative for an endemic steady state since $\beta \Omega > \gamma$. Hence the endemic fixed point is always asymptotically stable. Perturbations about the fixed point decay monotonically if the eigenvalues are purely real and in an oscillatory fashion if they are complex. These correspond, respectively, to overdamped and underdamped decay. In Figure (\ref{Over vs Underdamped decay}), we plot both time traces and phase portraits of S and I showing the underdamped and overdamped cases. Figure (\ref{fig:ratio_Im_Re_Eig}) is a state diagram of the model, showing the ratio $|\frac{Im({\lambda})}{Re({\lambda})}|$. The region of complex eigenvalues, corresponding to underdamped decay, is bounded by the contours labelled by $|\frac{Im({\lambda})}{Re({\lambda})}| = 0$.

\begin{figure}
\includegraphics[scale=0.4]{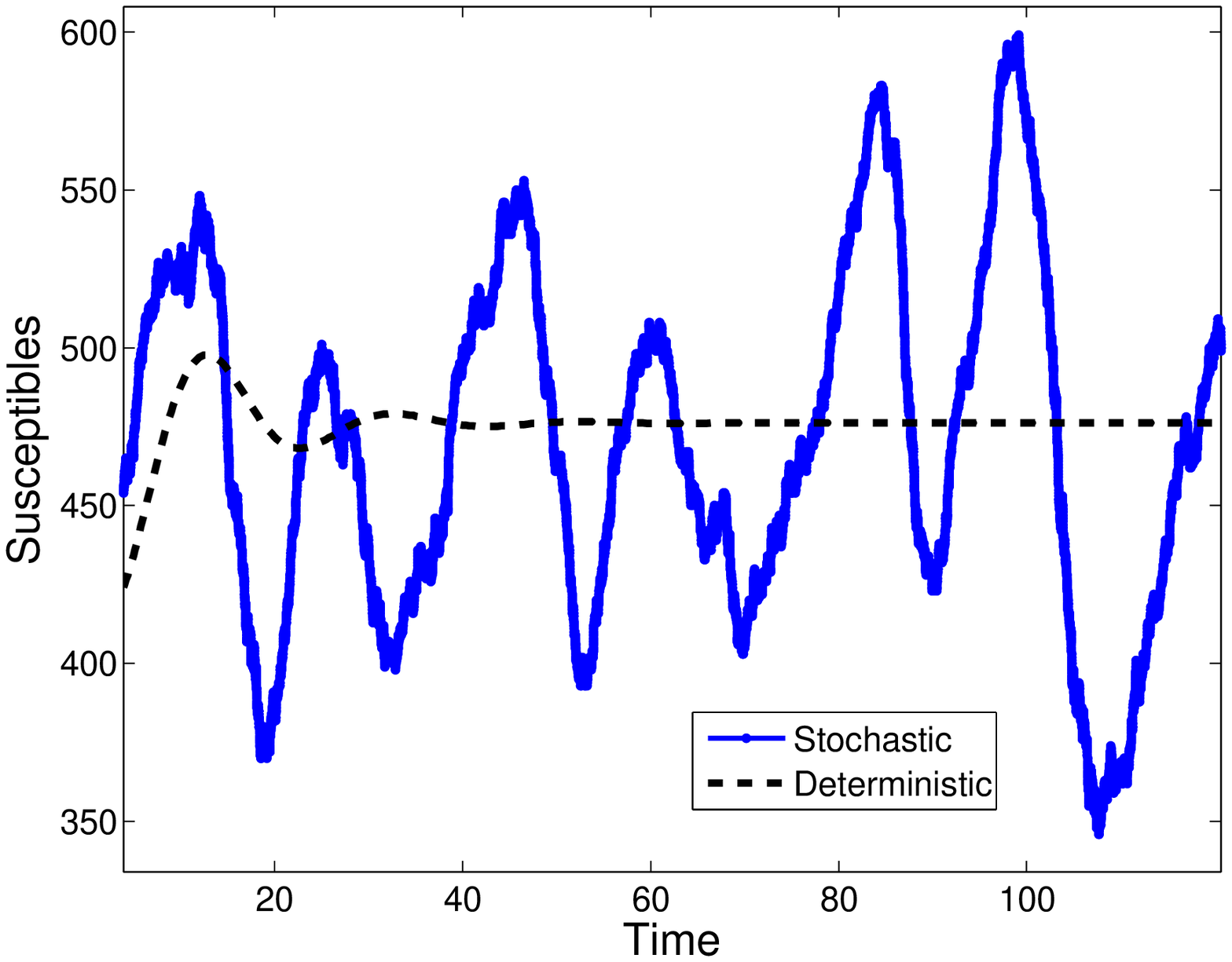}
\caption{\label{fig:stochastic_S}(Color online) Numerical simulation of susceptibles overlaid on deterministic underdamped decay. Parameters are $\beta=0.0021,\alpha=0.1, \gamma=1.0$ and population size $\Omega = 1000$. There are noise-induced oscillations in the stochastic case which are not seen in the deterministic analysis. The time period of oscillations is approximately $20$ in units of $1/\gamma$ (simulations have been performed after non-dimensionalisation). This corresponds well with the frequency seen (Figure \ref{fig:SIRS_PSD}) in the PSD analysis for the same set of parameter values.}
\end{figure}

\section{SIRS linear stochastic analysis}\label{Stochastic analysis}
 
Relative fluctuations about the deterministic expected values vary as the inverse of the square root of the number of interacting entities and thus become important when the entitites are few in number. Often one finds that this is indeed the case in biological systems \cite{schrodinger1992}. Our present study concerns populations where fluctuations due to demographic stochasticity cannot be ignored and mean-field deterministic analysis fails to capture its non-trivial contributions. It then becomes necessary to employ stochastic methods to reliably understand the role of fluctuations. 

We begin by writing down the birth-death master equation (ME) of the SIRS model. Let the state of the system at any time $t$ be given by the vector $\mathbf{n} = (n_1,n_2,n_3)$ where $n_i$ is the number of individuals in each class ($i=1$ for S, $i=2$ for I and $i=3$ for R). The general birth-death ME is then \cite{gardiner2004}

\begin{multline}
 \frac{\partial P(\mathbf{n},t)}{\partial t} = \sum_{\alpha} \left\{ t^-_{\alpha}(\mathbf{n}+\mathbf{r}_{\alpha}) P(\mathbf{n}+\mathbf{r}_{\alpha},t) - t^+_{\alpha}(\mathbf{n}) P(\mathbf{n},t) \right\} \\ 
 + \sum_{\alpha} \left\{ t^+_{\alpha}(\mathbf{n}-\mathbf{r}_{\alpha}) P(\mathbf{n}-\mathbf{r}_{\alpha},t) - t^-_{\alpha}(\mathbf{n}) P(\mathbf{n},t) \right\}
\end{multline}

Here $P(\mathbf{n},t)$ is the conditional probability for the system to be in the state $\mathbf{n}$ given some fixed initial state, $t^+_{\alpha}$ and $t^-_{\alpha}$ are the birth and death rate terms and $\mathbf{r}_{\alpha}$ is the vector denoting the change in the number of entities in the $\alpha$-th reaction. For the SIRS model we have
\begin{equation}
 \begin{split}
  t^+_1 &= \beta n_1 n_2; ~~ t^+_2 = \gamma n_2; ~~ t^+_3 = \alpha n_3; ~~ t^-_1 = t^-_2 = t^-_3 = 0 \\
  \mathbf{r}_1 &= (-1, +1,0); ~~ \mathbf{r}_2 = (0, -1, +1); ~~ \mathbf{r}_3 = (+1, 0, -1)
 \end{split}
\end{equation}
The variable $N(t) = n_1(t)+n_2(t)+n_3(t)$ is constrained by the fixed population size $\Omega$. Incorporating this constraint within the ME allows us to work with a $2 \times 2$ system. The partial time derivative of $P(N(t),t)$ vanishes if $N(t) = \Omega$ for all $t$. We marginalise with respect to one of the variables (here we choose $n_3$) taking the population size as parameter: $P(n_1, n_2, t\,|\, \Omega) = \sum_{n_3} \delta_{n_1 + n_2 + n_3, \, \Omega} \, P(\mathbf{n},t)$. Modifying the birth terms and the state change vectors appropriately, \emph{i.e.} replacing $n_3$ by $\Omega-n_1-n_2$ and writing the $\mathbf{r}_{\alpha}$ as $2 \times 1$ vectors, we get the marginalised ME.
\begin{multline}
   \frac{\partial P(n_1, n_2, t \,| \, \Omega)}{\partial t} = \beta (n_1+1) (n_2-1) P(n_1+1,n_2-1,t \,| \, \Omega) \\ + \gamma \, (n_2+1) \, P(n_1, n_2+1, t \,| \, \Omega) \\ + \alpha \, (\Omega-n_1-n_2+1) P(n_1-1, n_2, t \,| \, \Omega) \\ - \left\{ \beta n_1 n_2 + \gamma n_2 + \alpha \, (\Omega-n_1-n_2) \right\} P(n_1, n_2, t \,| \, \Omega)
\label{marginalised CME}
\end{multline}

The transition probability for the infection step is non-linear and as such the ME is not solvable analytically. However, it is possible to simulate the ME using the Doob-Gillespie stochastic simulation algorithm (SSA) \cite{doob1945, gillespie1976, gillespie1977}. This generates an exact sampled trajectory  of the jump stochastic process described by the ME. We non-dimensionalise time by working in units of $1/\gamma$. Figure (\ref{fig:stochastic_S}) shows the numerical simulation of the susceptibles using the SSA, compared with a deterministic solution of the ODE system. The demographic fluctuations induce and sustain approximate cycles in the populations, a feature absent in the deterministic model. 

In the absence of exact solutions, we try to characterise these fluctuations within an approximation method due to van Kampen \cite{vanKampen2007} which replaces the jump process with a stationary multivariate Ornstein-Uhlenbeck process. The Gaussian nature of the process can then be utilised to obtain analytical solutions for the fluctuation properties, while the linear nature of the process can be utilised to make connections between the deterministic and fluctuating dynamics.

We expand the variables in the population size $\Omega$ (the large parameter of the approximation method) so that the size of the jumps decreases as the population is increased, 
\begin{equation}
 \mathbf{n} = \Omega \, \overline{\mathbf{n}} + \Omega^{1/2} \mathbf{x}
 \label{van Kampen Omega expansion}
\end{equation}
where $\overline{\mathbf{n}}$ is the mean value of $\mathbf{n}$ and $\mathbf{x}$ denotes the fluctuations around the mean. 
Assuming the fluctuations obey a diffusion process about the mean yields a Fokker-Planck equation (FPE) for the fluctuations,
\begin{equation}
 \partial_t P(\mathbf{x},t) = -\partial_i \left[ A_i(\mathbf{x}) P(\mathbf{x},t) \right] + \frac{1}{2}  \partial_i \partial_j \left[ B_{ij}(\mathbf{x}) P(\mathbf{x},t) \right]
 \label{linearFPE}
\end{equation}
where repeated indices indicate summation, $\partial_t = \partial/\partial t$ and $\partial_i = \partial/\partial x_i$. This is the linear noise approximation. The elements of the drift vector $\mathbf{A(x)}$ and the diffusion matrix $\mathbf{B}(\mathbf{x})$ are given, following the prescription in Gardiner \cite{gardiner2004}, as
\begin{equation}
 \begin{split}
 A_i(\mathbf{x}) &= \sum_{\alpha=1}^2 r^i_\alpha t^+_\alpha(\mathbf{x}) \\
 B_{ij}(\mathbf{x}) &= \sum_{\alpha=1}^2 r^i_\alpha r^j_\alpha t^+_\alpha(\mathbf{x})
 \end{split}
\label{FPE A and B}
\end{equation}
 ($i,j=1,2$ being the component indices). Linearising a second time about the endemic fixed point we get the FPE of a stationary multivariate Ornstein-Uhlenbeck process
 \begin{equation}
  \partial_t P(\mathbf{x},t) = -\sum_{i,j} \left[ A_{ij} \partial_i \left\{x_j P(\mathbf{x},t) \right\} - \frac{1}{2} B_{ij} \partial_i \partial_j P(\mathbf{x},t)  \right] 
  \label{Linearised FPE}
 \end{equation}
where $A_{ij}$ and $B_{ij}$ are the elements of the linearised drift and diffusion matrices. For the SIRS model, their values are (from Equation (\ref{FPE A and B}) after putting $\mathbf{x}=\mathbf{x}^{\ast}$)

\begin{align}
 \mathbf{A} =
 \begin{bmatrix}
  - \alpha \left( \frac{\alpha+\beta \Omega}{\alpha + \gamma} \right) & -(\alpha+\gamma) \\
  \alpha \left( \frac{\beta \Omega -\gamma}{\alpha + \gamma} \right) & 0
 \end{bmatrix}
 \label{eq:SIRS_A}
\end{align}

\begin{align}
  \mathbf{B} = \frac{\alpha \gamma}{\beta} \left\{ \frac{\beta \Omega-\gamma}{\alpha+\gamma} \right\}
  \begin{bmatrix}
     2 & -1 \\
     -1 & 2
  \end{bmatrix}
\label{eq:SIRS_B}
\end{align}
We note that this linearised drift matrix $\mathbf{A}$ is identical to the linearised Jacobian matrix (Equation \ref{SIRS Linearised A matrix}) obtained from the deterministic analysis and hence the two matrices share the same spectrum. This allows us to predict, under the two-stage linearisation procedure, the existence of non-trivial stochastic phenomena like noise-induced quasicycles and stochastic coherence purely from a deterministic analysis of the spectral structure of the linearised Jacobian. We shall discuss this important point in greater detail in sections \ref{Quasicycles} and \ref{Stochastic coherence}. This also allows us to use the terms ``linearised drift matrix'' and ``linearised Jacobian matrix'' interchangeably.

The multivariate Ornstein-Uhlenbeck process has exact solutions for both stationary and transition probability densities. Both are multivariate Gaussians, fixed by the equal time covariance matrix $\Sigma_{ij} = \langle \langle x_i x_j \rangle \rangle$ and the matrix of time correlations $C_{ij}(\tau) = \langle \langle x_i(t) x_j(t+\tau) \rangle \rangle$, where the double angular brackets denote the cumulant \cite{vanKampen2007}. $\mathbf{\Sigma}$ can be obtained by solving the steady state Einstein relation \cite{gardiner2004, vanKampen2007}.
\begin{equation}
  \mathbf{A} \mathbf{\Sigma} + \mathbf{\Sigma}\mathbf{A}^T + \mathbf{B} = 0
  \label{eq:einstein_relation}
\end{equation}
This has the form of a matrix Lyapunov equation, and can be solved using a method first proposed by Barnett and Storey \cite{barnett1966} in the context of linear control systems. We note that Equation (\ref{eq:einstein_relation}) can be written as the sum of a matrix and its transpose $\mathbf{S} + \mathbf{S}^T = 0$ where $\mathbf{S}$ is the anti-symmetric matrix $ \mathbf{A}\mathbf{\Sigma} + \frac{1}{2} \mathbf{B}$. We can solve for $\mathbf{S}$ in terms of $\mathbf{A}$ and $\mathbf{B}$ using the relation 
\begin{equation}
  \mathbf{A}\mathbf{S} + \mathbf{S}\mathbf{A}^T = \frac{1}{2}\left( \mathbf{B}\mathbf{A}^T - \mathbf{A}\mathbf{B} \right)
  \label{eq:solve_for_S}
\end{equation}
which is obtained by eliminating $\mathbf{\Sigma}$ from the Einstein relation and using the definition of $\mathbf{S}$. Since $\mathbf{S}$ is anti-symmetric, it is specified by a single parameter when it is of size $2 \times 2$. This parameter can be obtained directly from Equation (\ref{eq:solve_for_S}), since both $\mathbf{A}$ and $\mathbf{B}$ are two-dimensional matrices and are known. For higher dimensions, matrix decompositions are convenient when solving for $\mathbf{S}$. 

For the SIRS model (using Equations \ref{eq:SIRS_A} and \ref{eq:SIRS_B}), we have
\begin{equation}
 \mathbf{S} = \left\{ \frac{\gamma(\beta\Omega-\gamma)(\alpha^2+2\alpha\gamma+2\gamma^2+\alpha\beta\Omega)}{2\beta(\alpha+\gamma)(\alpha+\beta\Omega)} \right\}
 \begin{bmatrix}
  0 & -1 \\ 1 & 0
 \end{bmatrix}
\label{SIRS S matrix}
\end{equation}
Knowing $\mathbf{S}$, $\mathbf{A}$ and $\mathbf{B}$ we can now write down the covariance matrix
\begin{equation}
 \mathbf{\Sigma} = \mathbf{A}^{-1} \left(\mathbf{S} - \frac{1}{2} \mathbf{B} \right)
 \label{NEQ covariance}
\end{equation}
which for the SIRS model is
\begin{equation}
 \mathbf{\Sigma} = \frac{\gamma}{\beta}
 \begin{bmatrix}
  \frac{\alpha^2+\gamma^2+\alpha(\beta\Omega+\gamma)}{\alpha(\alpha+\beta\Omega)} & -1 \\
  -1 & \frac{\alpha(\alpha+\beta\Omega)^2+\gamma(\alpha+\gamma)(\beta\Omega-\gamma)}{\beta(\alpha+\gamma)^2(\alpha+\beta\Omega)}
 \end{bmatrix}
 \label{SIRS Covariance}
\end{equation}
Having obtained the matrix $\mathbf{\Sigma}$, the matrix of time correlations follows as 
 \begin{equation}
 \mathbf{C}(\tau) = \langle \langle \mathbf{x}(t) \mathbf{x}(t+\tau) \rangle \rangle = e^{\tau \mathbf{A}} \mathbf{\Sigma}
 \label{eq:correlation_matrix}
\end{equation}
The stochastic SIRS model, in the linear noise approximation, is completely specified by $\mathbf{\Sigma}$ and $\mathbf{C}(\tau)$. In the next section we use quantities derived from these to examine the model for signatures of oscillatory behaviour.

\section{Noise-induced oscillations: Endogenous Quasicycles}\label{Quasicycles}
 
The trace of the variation of the populations with time shown in Figure (\ref{fig:stochastic_S}) is strongly suggestive of sustained oscillations. This can be verified quantitatively by measuring the power spectral density (PSD) of the population time series. A peak in the PSD indicates the presence of oscillations. The PSD matrix, in terms of the linearised drift and diffusion matrices for a multivariate Ornstein-Uhlenbeck process is
\begin{equation}
 \mathbf{P}(\omega) = (-i\omega \mathbb{I} + \mathbf{A})^{-1} \, \mathbf{B} \, (i\omega \mathbb{I} + \mathbf{A}^T)^{-1}
 \label{General PSD}
\end{equation}
where $\mathbb{I}$ is the identity matrix. The diagonal elements of this matrix give an estimate of the periodicity in the relevant variables (here S and I). The $\mathbf{P}_{ii}$ for the SIRS PSD are
\begin{equation}
 \mathbf{P}_{ii}(\omega) = 2 d  \left( \frac{\Gamma_i + \omega^2}{\omega^4 + q\omega^2 + r} \right)
 \label{eq:P_ii}
\end{equation}
where $d = \frac{\alpha \gamma (\beta \Omega - \gamma)}{\beta (\alpha+\gamma)}$, $\Gamma_1 = (\alpha+\gamma)^2$, $\Gamma_2 = \alpha^2 \left\{ \alpha^2 + \gamma^2 + \beta \Omega(\beta \Omega - \gamma) + \alpha (\beta \Omega + \gamma) \right\} /(\alpha+\gamma)^2$, $q = \alpha \left\{ \alpha^2(\alpha+2\gamma) - 2 \gamma^2(\beta \Omega - \gamma) + \alpha(\beta \Omega - 2\gamma)^2 \right\}/(\alpha+\gamma)^2$ and $r = \alpha^2(\beta \Omega -\gamma)^2$. 
 
\begin{figure}
\includegraphics[scale=0.4]{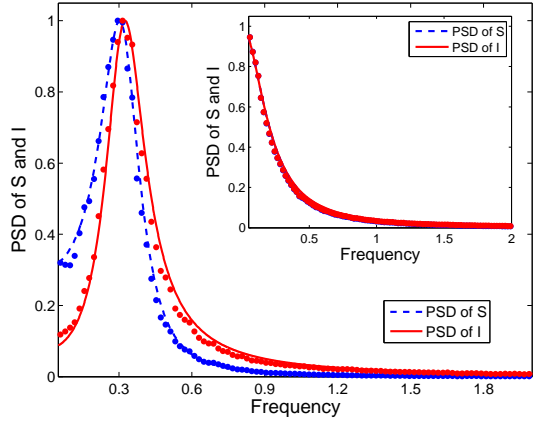}
\caption{\label{fig:SIRS_PSD}(Color online) Normalised power spectral density for S and I. There is excellent agreement between the analytically calculated (dashed blue for S, solid red for I) and numerically computed (blue dots for S, red dots for I) PSDs, thus justifying the linear noise approximation. The main graph is for parameter values $\beta=0.0021, \alpha = 0.1, \gamma = 1.0$ at population size $\Omega = 1000$ which falls within the underdamped zone. The PSD peaks around frequency $\omega=0.3$, which corresponds approximately to the time period of the numerical signal as well as that of the ACF for the same set of parameter values (see Figures \ref{fig:stochastic_S} and \ref{fig:SIRS_ACF}). The inset shows the PSD for the same size of the population at parameter values $\beta=0.0012,\alpha = 2.5, \gamma=1.0$ which falls within the overdamped zone and does not show any peak.}
\end{figure}

In Figure (\ref{fig:SIRS_PSD}) we plot the PSD for both S and I, comparing numerical simulation with Equation (\ref{eq:P_ii}). A peak is clearly visible for parameters corresponding to underdamped dynamics. The peak disappears for overdamped dynamics as shown in the inset. The peak frequency (around $\omega_p=0.3$) corresponds to the period ($T = 20$) of the numerical time-trace (Figure \ref{fig:stochastic_S}). The excellent agreement between numerics and analytics provides a post-facto justification of the linear noise approximation for this problem.

The PSD has peaks at real frequencies if and only if the extremum condition $\partial P_{ii}(\omega) / \partial \omega = 0$ has real roots. The regions of parameter space for which this occurs are bounded by contours labelled ``$\partial P_{11}(\omega) / \partial \omega = 0$'' and ``$\partial P_{22}(\omega) / \partial \omega = 0$'' in Figure (\ref{fig:ratio_Im_Re_Eig}). These are sufficient conditions for the existence of quasicycles. This approach has been used previously in the literature to detect quasicycles \cite{mckane2005, alonso2007, rozhnova2009}.

\begin{figure}
\includegraphics[scale=0.4]{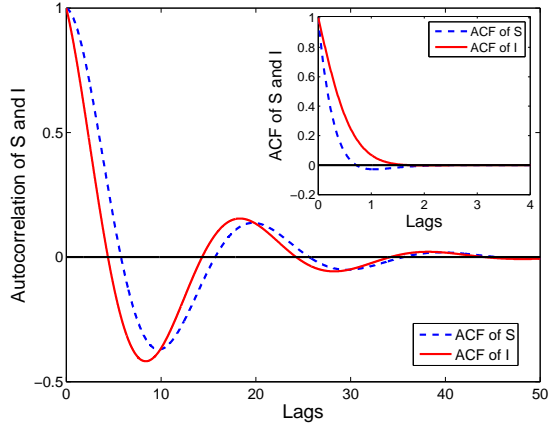}
\caption{\label{fig:SIRS_ACF}(Color online) Normalised autocorrelation of susceptibles and infected for parameter values $\beta=0.0021,\alpha = 0.1, \gamma=1.0$ and population size $\Omega = 1000$ which falls within the $|\frac{Im({\lambda})}{Re({\lambda})}| > 1$ zone. The thick black line is the x-axis. There is clear oscillatory decay with a period of approximately $20$ in units of $1/\gamma$ which agrees well with Figures (\ref{fig:stochastic_S}) and (\ref{fig:SIRS_PSD}) for the same set of parameter values. The inset plot shows the ACF for parameter values $\beta=0.009,\alpha = 1.2, \gamma=1.0$ and population size $\Omega = 1000$ which falls within the zone bounded by the contours $\partial P_{11}(\omega) / \partial \omega = 0$, $\partial P_{22}(\omega) / \partial \omega = 0$ and $|\frac{Im({\lambda})}{Re({\lambda})}| = 0$, \emph{i.e} the region where susceptibles should cycle according to the PSD analysis. The thick black line is once again the x-axis. There is a single zero-crossing of the ACF for S, which indicates non-oscillatory decay \cite{nisbet1982}. This shows the unreliability of PSD analysis in detecting quasicycles. These analytical plots have also been compared with numerical data (not shown here) with good agreement.}
\end{figure}

While Fourier analysis of a signal is a natural tool for studying oscillatory behaviour, a corresponding time-domain analysis must yield equivalent results. The time-correlation function forms the basis of a time-domain analysis, which for the multivariate Ornstein-Uhlenbeck process is given by Equation (\ref{eq:correlation_matrix}). The temporal variation of the time correlation is fixed entirely by the drift $\mathbf{A}$ which is the deterministic part of the dynamics, while its scale is set by $\mathbf{\Sigma}$ which involves the stochastic part of the dynamics through $\mathbf{B}$. Defining a normalised time correlation $\mathbf{c}(\tau) = \mathbf{C}(\tau) \mathbf{\Sigma}^{-1}$, we find that $\mathbf{c}(\tau) = e^{\tau \mathbf{A}}$. This is of the form $c(\tau) \sim \exp[Re(\lambda) \tau] \sin[Im(\lambda) \tau)]$. This observation motivates the use of the ratio $|\frac{Im({\lambda})}{Re({\lambda})}|$ to reliably detect quasicycles within the linear noise approximation, where $\lambda=\mathrm{eig}(\mathbf{A})$. If the decay time scale, fixed by Re($\lambda$), is too short compared to the oscillatory time scale fixed by Im($\lambda$), the decay will dominate and oscillatory effects will not be discernible. This will be so even when the extremum condition has real roots. We thus propose a condition for clearly discernible quasicycles, namely $|\frac{Im({\lambda})}{Re({\lambda})}| \geq 1$. In Figure (\ref{fig:ratio_Im_Re_Eig}) we plot the contour $|\frac{Im({\lambda})}{Re({\lambda})}| = 1$. The region $|\frac{Im({\lambda})}{Re({\lambda})}| > 1$ is bounded on the right by this contour. As this is more stringent than the extremum condition $\partial P_{ii}(\omega) / \partial \omega = 0$, it is entirely contained by the regions where the PSD has a peak. In Figure (\ref{fig:SIRS_ACF}) we emphasise this point by comparing the ACF when the PSD has peaks at finite frequencies. When $|\frac{Im({\lambda})}{Re({\lambda})}|$ is small the oscillations are barely discernible as seen from the rapid decay of the ACF. For $|\frac{Im({\lambda})}{Re({\lambda})}|$ of order unity clear signatures of oscillation are visible in the ACF.

\section{Quality of noise-induced oscillations: Stochastic coherence}\label{Stochastic coherence}

Noise-induced oscillations, unlike genuine oscillations, are not phase coherent and as such are called quasicycles. The coherence or regularity of quasicycles can be quantified by several measures. Here we use the quality factor, which measures the sharpness of the peak of the PSD, and the coefficient of variation which measures the regularity of the zero crossing of the signals.

\begin{figure}
\subfigure[Quality Factor]
{\includegraphics[scale=0.4]{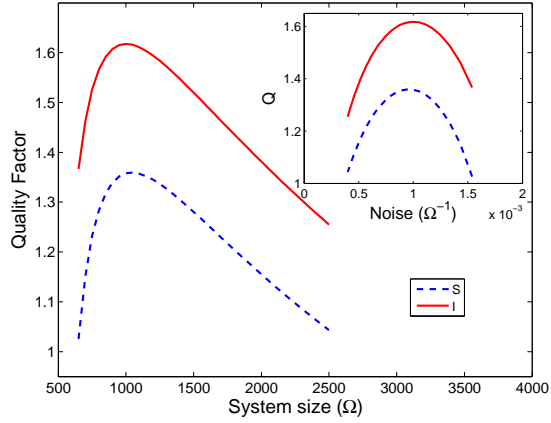}
\label{Quality factor for S and I}} \\
\subfigure[Coefficient of Variation]
{\includegraphics[scale=0.37]{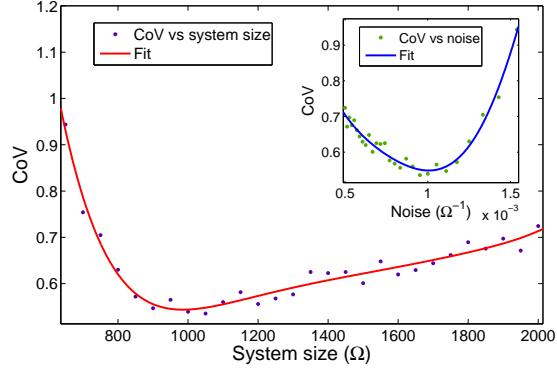}\label{CoV for S}}
\caption{(Color online) (a) Quality factor and (b) Coefficient of Variation for the susceptibles (plotted for $200$ runs) against population size ($\Omega$) and (inset) against inverse of population size. The solid line in (b) is a fifth-order polynomial fit. Parameter values are $\beta=0.0021,\alpha = 0.1, \gamma=1.0$ for (a) and $\beta=0.002,\alpha = 0.1, \gamma=1.0$ for (b). In the $Q$ plot, the peaks for $S$ and $I$ are respectively at $\Omega=1040$ and $\Omega=1000$, while the $C_V$ peak for $S$, computed from the minimum of the fifth-order polynomial fit, is at $\Omega = 985$. The $C_V$ plot for the infected (not given here) shows a peak (in the sixth-order fit used there) at $\Omega=1135$.}
\end{figure}

The quality factor, $Q$, is a dimensionless parameter that characterizes an oscillator's bandwidth relative to its peak frequency, 
\begin{equation}
 Q = \omega_p/\Delta \omega
 \label{General quality factor}
\end{equation}
where $\omega_p$ is the peak frequency and $\Delta \omega$ is the bandwidth. A high $Q$ corresponds to oscillations of greater regularity. We calculate the $Q$ for the diagonal entries of the PSD matrix. Let $k_i$ be half the maximal power $k_i = \frac{1}{2} P_{ii}(\omega_p^i)$ for each $i=1,2$. We calculate the bandwidth or the full-width at half-maximum (FWHM) using the $k_i$ and Equation (\ref{eq:P_ii}) to get 
\begin{equation}
(\Delta \omega)_i = \sqrt{(2d/k_i-q)-2\sqrt{r-2d \Gamma_i/k_i}}
\label{FWHM}
\end{equation}
Using Equation (\ref{eq:P_ii}), the peak frequencies are given by the positive square roots of the positive roots of the two quadratic equations  $z^2 + 2\Gamma_i z + (\Gamma_i q - r) = 0$ (for $i=1,2$) where $z=\omega^2$ and $\Gamma_i$, $q$ and $r$ are as defined in the previous section. The peak frequency and the FWHM together give the $Q$. Figure \ref{Quality factor for S and I} shows a scan of the quality factor against population size and against the inverse of population size (inset). As one would expect, $Q$ is low for high noise amplitudes and starts increasing as the noise decreases, keeping in mind that the relative noise amplitude varies as the inverse of the square root of the size of the population. However, the graph then has a maximum and then decreases for high amplitudes of noise. This is stochastic coherence.

The coefficient of variation, $C_V$, is the variance over the mean of the times $T$ between succesive zeros of a temporal signal. A sharp peak in the histogram of the intervals between zero crossings, then, indicates a strongly coherent signal. $C_V$ is a dimensionless measure of this, 
\begin{equation}
 C_V=\frac{\sqrt{\langle T^2 \rangle - \langle T \rangle^2}}{\langle T \rangle}
 \label{Coefficient of variation}
\end{equation}
A low $C_V$ indicates a high degree of coherence in the signal. Similar measures are used in the literature (for example \cite{pikovsky1997} and \cite{hilborn2008}). Figure \ref{CoV for S} shows the $C_V$ for the mean crossing time of the numerical signal of the susceptibles scanned against population size and (inset) against its inverse . The plot shows a minimum which indicates stochastic coherence and hence numerically supports the analytical result given by the $Q$.

Although this non-intuitive variation of the coherence of the quasicyles with the size of the population has a stochastic origin, it is controlled purely by the deterministic part of the dynamics. The analysis using the $Q$ and the $C_V$ require a knowledge of the diffusion matrix $\mathbf{B}$. However, after the two-step linearisation procedure, the entire non-trivial dependance on the population size is contained only in the spectrum of the linearised drift matrix, while the diffusion matrix scales linearly with $\Omega$, as given by Equations (\ref{eq:SIRS_A}) and (\ref{eq:SIRS_B}). Thus, any non-monotonicities in the fluctuations arise purely from the deterministic part of the dynamics, while the noise merely excites these modes. For a system which can be reduced to a standard multivariate Ornstein-Uhlenbeck process, the linearised drift matrix is identical to the linearised Jacobian matrix. This motivates the use of the ratio $|\frac{Im(\lambda)}{Re(\lambda)}|$ in determining the size of the population at which stochastic coherence is observed. This allows us to study stochastic coherence from the deterministic part of the dynamics. 

\begin{figure}
\includegraphics[scale=0.4]{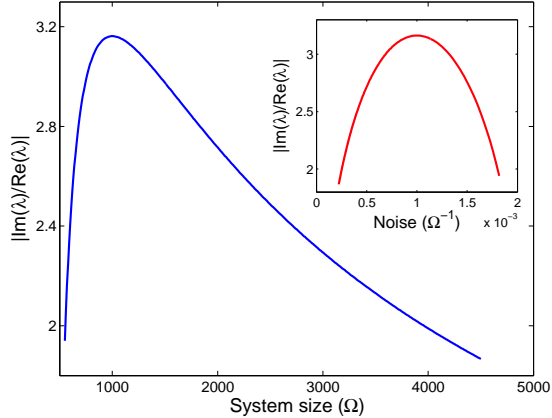}
\caption{\label{fig:ratio_Im_Re_Eig_scan}(Color online) Absolute value of the ratio of the imaginary and real parts of the eigenvalues of the linearised Jacobian matrix plotted against population size ($\Omega$) and (inset) against inverse of population size. Parameter values are $\beta=0.0021,\alpha = 0.1, \gamma=1.0$. There is a peak at $\Omega=1000$ corresponding to the stochastic coherence point.}
\end{figure}

We observe that in Figure (\ref{fig:ratio_Im_Re_Eig_scan}) the ratio $|\frac{Im(\lambda)}{Re(\lambda)}|$, when scanned against the size of the population ($\Omega$), shows a peak which occurs at $\Omega=1000$ for the parameters $\beta=0.0021$, $\alpha=0.1, \gamma = 1.0$.  We see that this value matches well with the peaks in Figures \ref{Quality factor for S and I} and \ref{CoV for S}, within numerical errors. We have calculated the peak value of the ratio in terms of the model parameters. If $\Omega_p$ is the population size at stochastic coherence, then 
\begin{equation}
 \Omega_p = \frac{\alpha+2 \gamma}{\beta}
 \label{Stochastic coherence system size}
\end{equation}
Since the ratio is always positive, there is stochastic coherence for all values of parameters for which quasicycles exist.

\section{Detailed balance violation necessary for quasicycles}\label{Non-equilibrium}

Quasicyles and stochastic coherence are not possible unless detailed balance is violated in the master equation. Typically, variables characterising biological systems are even under time-reversal, $x(-t) = x(t)$, and this is true for the variables of the SIRS model. If the system is in state $\mathbf{n}_0$ at time $t_0 = 0$ (say) and is in state $\mathbf{n}$ at some later time $t$, then the joint probability of the forward transition ($\mathbf{n}_0(t_0) \rightarrow \mathbf{n}(t)$) is $P(\mathbf{n},t;\mathbf{n}_0, 0)$ while that of the reverse ($\mathbf{n}(t_0) \rightarrow \mathbf{n}_0(t)$) is $P(\mathbf{n}_0,t;\mathbf{n}, 0)$ provided all time-reversal parities are even. Microscopic reversibility implies that at equilibrium the steady state forward and reverse joint probabilities must be equal. This is the condition for detailed balance and for a Markov process can be written as
\begin{equation}
 P(\mathbf{n},t | \mathbf{n}_0, 0) P_s (\mathbf{n}_0) = P(\mathbf{n}_0,t | \mathbf{n}, 0) P_s (\mathbf{n})
 \label{Microreversibility}
\end{equation}
where the subscript `s' denotes steady state. Expressing this condition macroscopically in terms of the correlation function, expanding in Taylor series, and keeping the first order terms one obtains \cite{lax1960} the Onsager relations
\begin{equation}
 \mathbf{A} \mathbf{\Sigma} = \mathbf{\Sigma} \mathbf{A}^T
 \label{Onsager relations}
\end{equation}
which is the macroscopic condition for equilibrium. This condition requires the drift matrix to be related by a similarity transformation to a symmetric matrix \cite{gardiner2004} and hence restricts its spectrum to the real axis. Since it is not possible to have quasicycles without having complex eigenvalues, the violation of detailed balance is a necessary condition for the existence of noise-induced oscillations. 

Recalling that $\mathbf{S} = \mathbf{A} \mathbf{\Sigma} + \frac{1}{2} \mathbf{B}$ and using the symmetry properties of $\mathbf{\Sigma}$ and $\mathbf{B}$ we can write down the following expression for $\mathbf{S}$ \cite{tomita1974}
\begin{equation}
 \mathbf{S} = \frac{1}{2} (\mathbf{A} \mathbf{\Sigma} - \mathbf{\Sigma} \mathbf{A}^T)
 \label{S as deviation from detailed balance}
\end{equation}
which then is a measure of the deviation from detailed balance. The SIRS $\mathbf{S}$ matrix (Equation \ref{SIRS S matrix}) can never be zero for any choice of parameters under the endemic condition $\beta \Omega > \gamma$. Thus the SIRS model always violates detailed balance and therefore allows for quasicycles for any choice of parameters.

\section{Non-normality increases Variance}\label{Non-normality}

We have already noted that the violation of detailed balance is necessary for quasicycles. Here we further note that detailed balance violation has another consequence, that of enhancement of fluctuation amplitudes. With detailed balance the drift matrix $\mathbf{A}$ is similar to a symmetric matrix, and is therefore normal ($\mathbf{A A}^T = \mathbf{A}^T \mathbf{A}$). In the absence of detailed balance, the drift matrix is no longer symmetric, and in this case is also non-normal. 

As has been noted by Ioannou \cite{ioannou1995}, the variance of a non-normal system driven by diagonal white noise is larger than its normal counterpart. Consider two stationary multivariate Ornstein-Uhlenbeck processes with drift and diffusion matrices ($\mathbf{A}_1,\mathbf{B}$) and ($\mathbf{A}_2,\mathbf{B}$) where $\mathbf{A}_1$ is non-normal but shares the same eigenvalues as the normal $\mathbf{A}_2$. Then, Schur decompositions of the two matrices gives $\mathbf{A}_1 = \mathbf{U} (\mathbf{D}+\mathbf{T}) \mathbf{U}^{\dagger}$ and $\mathbf{A}_2 = \mathbf{U} \mathbf{D} \mathbf{U}^{\dagger}$ where $\mathbf{U}$ is unitary, $\mathbf{D}$ is diagonal matrix of eigenvalues and $\mathbf{T}$ is strictly upper triangular. Restricting the forcing to be diagonally correlated white noise ($\mathbf{B}=\mathbb{I}$), Ioannou shows that ${\rm Tr}(\mathbf{\Sigma}_1 ) \geq {\rm Tr}(\mathbf{\Sigma}_2)$, where $\mathbf{\Sigma}_1$ and $\mathbf{\Sigma}_2$ are the respective covariance matrices and ${\rm Tr}(\ldots)$ denotes the trace of a matrix. For a general $\mathbf{B}$ which is not necessarily diagonal, $\mathbf{\Sigma}_1 = \mathbf{A}_1^{-1} \left( \mathbf{S}-\frac{1}{2}\mathbf{B} \right)$ and $\mathbf{\Sigma}_2 = -\frac{1}{2} \mathbf{A}_2^{-1} \mathbf{B}$. We have calculated the ratio of the traces of $\mathbf{\Sigma}_1$ and $\mathbf{\Sigma}_2$.

\begin{equation}
 \frac{{\rm Tr}(\mathbf{\Sigma}_1)}{{\rm Tr}(\mathbf{\Sigma}_2)} = 1+\frac{{\rm Tr}\left( \mathbf{A}_1^{-1} \mathbf{S} + \frac{1}{2 \Delta} \mathbf{UTU}^{\dagger}\mathbf{B} \right)}{{\rm Tr}\left( \mathbf{\Sigma}_2 \right)}
 \label{Ratio of traces of covariance matrices}
\end{equation}
where $\Delta$ is the determinant of $\mathbf{A}_1$. This expression is valid only when the spectrum of $\mathbf{A}_1$ is purely real. For the SIRS model, this ratio is greater than unity.

\begin{figure}
 \includegraphics[scale=0.5]{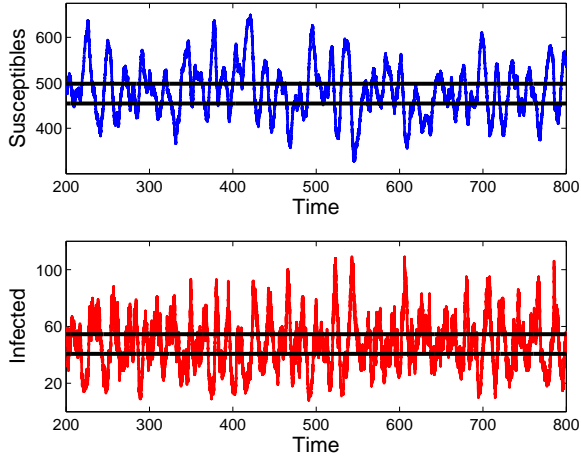}
 \caption{\label{Enhanced variance of time-traces of S and I}(Color online) Enhanced variance of time-traces of susceptibles and infected for parameter values $\beta=0.0021,\alpha = 0.1, \gamma=1.0$ with population size $\Omega=1000$. The solid black lines correspond to $\overline{n} \pm \sqrt{\overline{n}}$, where $\overline{n}$ is the mean.}
 
\end{figure}

Individual time-traces also show an increase in variance. Figure (\ref{Enhanced variance of time-traces of S and I}) shows time-traces of S and I where the fluctuations are seen to be  higher than the expected standard deviation values ($\overline{n} \pm \sqrt{\overline{n}}$, where $\overline{n}$ is the mean) marked by the black lines. 

\section{Conclusion}

In this paper, we have analysed a closed endemic model for sustained, though asymptotically incoherent, oscillations in the population classes. These oscillations are generated through fluctuations brought about by internal demographic stochasticity which destabilise the endemic fixed point. The closed nature of the problem allows one to deal with a simplified lower-dimensional problem, an aspect we have exploited systematically by showing how the master equation can be marginalised using the constraint. This model also lends itself to a two-stage linearisation procedure, at the end of which it is reduced to a multivariate Ornstein-Uhlenbeck form. This results in the identification of the linearised drift matrix with the deterministic Jacobian matrix linearised about the endemic fixed point and permits the analysis of stochastic behaviour from the deterministic behaviour.

Noise-induced oscillations or quasicycles are possible only if the eigenvalues of the linearised Jacobian matrix are complex. These oscillations are distinct from those produced by external periodic agencies because their phases decorrelate asymptotically. Quasicycles can be reliably detected only if the oscillation time period is at least of the same order as the decorrelation time scale, as otherwise the decay dominates over the oscillation. Strong quasicycles are seen when the imaginary parts of the eigenvalues are larger than the real parts.

Stochastic coherence, or the non-trivial maximisation of regularity of the oscillations at intermediate relative noise amplitudes (or equivalently at intermediate population sizes), is a striking aspect of the SIRS quasicycles. We have seen this both analytically from the relative strength of the peak of the power spectral density and numerically by directly computing the signal-to-noise ratio of the time-traces of each population class. This analysis requires a knowledge of the intrinsic noise in the system, namely the diffusion matrix $\mathbf{B}$. However, we find that, for systems which can be reduced to a standard multivariate Ornstein-Uhlenbeck form by the two-stage linearisation procedure mentioned earlier, it is possible to predict stochastic coherence purely from the deterministic analysis. Any non-trivial dependance on population size is contained only in the eigenvalues of the linearised drift matrix or equivalently the linearised Jacobian matrix, while the diffusion matrix scales linearly with the population size. Thus, any non-monotonicities in fluctuations arise entirely from the deterministic part of the dynamics, i.e. the spectrum of the drift matrix. The noise merely excites these modes. This motivates the maximisation of the ratio $|\frac{Im{(\lambda)}}{Re{(\lambda)}}|$ in the investigation of the population size value at which stochastic coherence is observed. Numerical results support this procedure. Therefore, we conclude that it is possible to make predictions about non-trivial behaviour of such systems in the stochastic regime by simply analysing the linearised deterministic dynamics.

The violation of detailed balance is a necessary condition for the existence of quasicycles. Analysis of the drift, diffusion and $\mathbf{S}$ matrices indicates that the population system described by the SIRS model is always out of equilibrium and allows for quasicycles about the endemic fixed point for any choice of model parameters. Violation of detailed balance due to the non-normal nature of the system dynamics is  manifest in the enhancement of fluctuation amplitudes of the populations. We have given an expression for the ratio of the trace of the non-normal covariance matrix over its normal counterpart, restricted to parameter values where the Jacobian spectrum is purely real. Numerics indicate that this ratio is greater than unity for the SIRS model.

The analysis of this paper shows that the phenomenon of noise-induced oscillations and stochastic coherence can generically be expected in non-equilibrium birth-death jump Markov processes which can be reduced to the standard multivariate Ornstein-Uhlenbeck form by a successive application of two linearisation procedures: the linear noise approximation followed by a linearisation about the fixed point of the system. This may therefore explain the appearance of asymptotically incoherent oscilations in other systems described by such equations, as for instance, in the repressilator \cite{rep2010}.

\begin{acknowledgments}
 We thank Prof. Rama Govindarajan for bringing non-normality to our attention and Sandeep K Goyal for discussions. We thank Profs. I. Bose and G. I. Menon for a critical reading of the manuscript. We thank PRISM, DAE for funding.
\end{acknowledgments}

\bibliography{epi_bib}

\end{document}